**Disentangling transport mechanisms in a correlated oxide by photoinduced charge injection**

Henry Navarro[1]*, Sarmistha Das[1]*, Felipe Torres[2,3]*, Rourav Basak[1], Erbin Qiu[1], Nicolas M. Vargas[1,4], Pavel Lapa[1,4], Ivan K. Schuller[1†] and Alex Frano[1††]

[1] Department of Physics, Center for Advanced Nanoscience, University of California, San Diego, 92093, USA.
[2] Department of Physics, Universidad de Chile, Santiago 7800024, Chile.
[3] Center for the Development of Nanoscience and Nanotechnology, CEDENNA, Santiago 9170124, Chile.
[4] General Atomics, PO Box 85608, San Diego, CA 92186, USA

*These authors contributed equally

**Corresponding Author**

†E-mail: ischuller@ucsd.edu
††E-mail: afrano@ucsd.edu

We present a novel heterostructured approach to disentangle the mechanism of electrical transport of the strongly correlated $PrNiO_3$, by placing the nickelate under the photoconductor CdS. This enables the injection of carriers into $PrNiO_3$ in a controlled way, which can be used to interrogate its intrinsic transport mechanism. We find a non-volatile resistance decrease when illuminating the system at temperatures below the $PrNiO_3$ metal-insulator transition. The photoinduced change becomes more volatile as the temperature increases. These data help understand the intrinsic transport properties of the nickelate-CdS bilayer. Together with data from a bare $PrNiO_3$ film, we find that the transport mechanism includes a combination of mechanisms including both thermal activation and variable range hopping. At low temperatures without photoinduced carriers the transport is governed by hopping, while at higher temperatures and intense illumination the activation mechanism becomes relevant. This work shows a new way to optically control the low-temperature resistance of $PrNiO_3$.

The electrical transport of transition metal oxides is of widespread interest because a single material can display different properties, such as metal-insulator transitions (MITs), which can be tuned by external perturbations upon slight perturbations[1-17]. Because of this tunability, focus is placed on the microscopic origin of the MITs, believing that understanding the microscopic mechanism will enable control of the electronic properties. However, for many systems, the understanding of the temperature dependent resistance R(T) is largely unexplored, ill-understood, and frequently not discussed in the literature. An example of such a material is the rare-earth nickelates family ($R$NiO$_3$, $R$ = Rare-earth)–a quintessential correlated oxide that displays an MIT. Since the transition is intertwined with magnetic, structural and bond order, the origin of this transition has been a topic of intriguing debate. Ideas range from Fermi surface nesting[18, 19] to a charge or bond disproportionation[20-22] and others[23]. For $R$ = Pr or Nd, a first-order MIT occurs with a paramagnetic-to-antiferromagnetic phase transition at the same temperature ($T_{MI}$)[9, 24, 25]. The transition exhibits a pronounced thermal hysteresis, typical of first-order transitions, because of the coexistence of insulating and metallic phases. Below the hysteresis, the resistance increases exponentially with decreasing temperature. Despite this behavior seen frequently in samples from different research groups[18, 26-34], the mechanism of the transport below the transition remains an open question[19-21, 35-37]. Yet, interest in tuning the transport in these materials continues to grow, fueled by potential applications[28, 30, 38]. Possibly the lack of understanding of the transport is related to the effect of defects and sample-dependent details that do not carry over reliably from sample to sample. A methodology to investigate the transport of a single sample under variable controlled conditions would alleviate this difficulty.

Here we outline a method to interrogate the transport properties of PrNiO$_3$ (PNO), comparing transport regimes in a single sample. We designed a heterostructure that allows us to inject carriers and thus affect its transport properties in controlled way[39, 40]. We place a thin film of PNO in proximity to the photoconducting cadmium sulfide (CdS), which acts a source of photoinduced charge carriers[41]. While PNO does not have any response to visible light, when the heterostructure is illuminated with white light its low-temperature resistance drops by a factor of 10 allowing to explore intrinsic transport properties of insulating phase of the nickelate. This effect only manifests below the hysteretic regime of the first-order MIT, tying its relevance to the intrinsic ground state properties of PNO. The phenomenology was explored experimentally as a function of light intensity which is proportional to the number of carriers induced into the nickelate layer. Moreover, the heterostructure allows exploration of the temporal dynamics of the carriers, which are found to be long-lasting, or non-volatile, at low temperatures. As the temperature increases and approaches the hysteresis range, the photoinduced changes continually become more volatile.

Armed with this set of experimental observations, we use a model that describes the heterostructure's temperature-dependent resistance, with and without the effect of light, as well the resistance of a bare PNO film. The model combines thermally activated transport with variable range hopping in the nickelate layer, like in previous models[36, 37], but here we add the mechanism of photoinduced transport. A single model individually is insufficient, but rather a combination of the two is necessary to explain the data. At low temperatures and without light, the transport is mostly governed by hopping, while at high temperatures the thermal energy enhances the activated transport. The model then assumes that charge carriers migrate from the CdS layer into the nickelate when the light is turned on. The dependence of the resistive switching as a function of light intensity implies that at low intensity the transport is mostly via hopping, while at high intensity the transport is mostly activated. In addition, the model also explains the non-volatile time dependence at low temperatures and the volatile response at higher temperatures, by introducing a time-dependent carrier density without any additional adjustable parameters.

Our method provides a practical tuning knob and insights into the physics of nickelates. We have developed optoelectronic control of a nickelate's resistance, which adds to the list of methodologies used to gain control of the material's resistance such as doping, gating, pressure, magnetic and electric fields[9, 24, 42-45]. This may lead to new multimodal or hybrid functionalities. In addition, in the specific case of PNO, we find that the transport is governed by both thermally activated transport and variable hopping into impurity states near the bottom of the conduction band. This is consistent with other models[18, 26-28, 35-37], but in our case the interplay between these two effects can be controlled by injecting charge carriers into the oxide, achieved through an adjacent photoconducting layer.

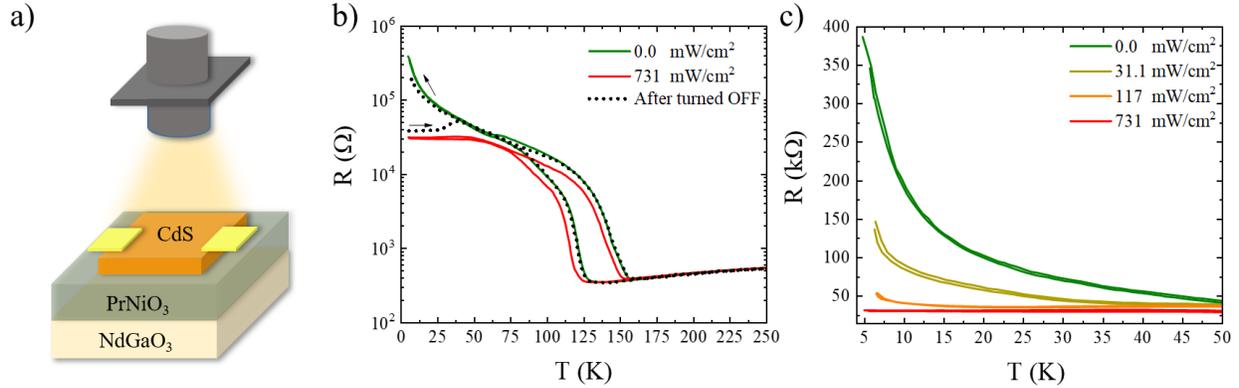

**FIG. 1.** (a) Experimental setup for the two terminal (Au electrodes) resistivity and photoconductivity of the CdS–PrNiO$_3$ heterostructure. During the electrical transport measurements, a white light emitting diode was positioned over the sample. (b) Temperature dependence of the sample resistance R(T) without light (green), and with light of 731 mW/cm$^2$ power density turned on (orange). The black dashed curve corresponds to the R(T) measured in warming from 7 K after the LED is turned off. (c) R(T) in the insulating state below 50 K, the temperature at which the hysteresis curves merge, measured for various light power densities.

The hybrid device (Fig. 1(a)) was fabricated by heterostructuring PNO with a photoconducting semiconductor, CdS. An epitaxial PrNiO$_3$ (PNO) 20-nm thick film on a single crystal NdGaO$_3$ (NGO) ⟨101⟩ substrate using pulsed laser deposition. On top of this layer, an 80-nm thick CdS film was deposited using rf magnetron sputtering. Two Au (40 nm) electrodes were patterned on top of the CdS–PNO heterostructures. X-ray diffraction measurements were performed in a Rigaku SmartLab system at room temperature, confirming the single-phase epitaxial growth along the ⟨101⟩ direction for PNO and a polycrystalline CdS (see Supplemental Material for Figure S1 diffraction patterns). Transport measurements were carried out on a Montana C2 S50, using a Keithley 6221 current source and a Keithley 2182A nanovoltmeter. A light emitting diode (LED) with variable power density was placed over the sample illuminating its surface (Figure 1(a)). In this geometry, light strongly affected the CdS and subsequently affected the bottom PNO layer through proximity effects. Since the resistance of bare CdS with light on is 2.5 orders of magnitude higher than the CdS–PNO bilayer (see Supplemental Material for Figure S3 electrical transport measurements), we consider the measured transport is through the nickelate.

Figure 1 summarizes one central finding of our work: the observation of a light-induced resistive switching in nickelates. Figure 1(b) shows the changes in the resistance versus temperature (R(T)) of the CdS–PNO hybrid when illuminated by the LED. Without illumination (green curves in Figure 1(b)), the CdS–PNO hybrid exhibits a first-order MIT at around 137 K, as has been commonly observed[31, 46-49]. The resistance shows a change of ~3 orders of magnitude near the hysteretic phase transition, and has a nonlinear temperature dependence in the insulating state (7–50 K). Upon illumination of the heterostructure with 731 mW/cm$^2$ white light (orange curve), the 7 K resistance drops by a factor of ~10. The downwards shift of the entire R(T) curve indicates a slight (~6K) heating by the light. However, this thermal shift is not large enough to account for the light-induced change in the low temperature (7 K) resistance. The resistance change would be equivalent to a temperature increase from 7 K to 50 K.

After turning off the white LED (dashed lines), the resistance of PNO increases slightly in the insulating state (7–50 K). When heating after turning off the light, the resistance recovers its original value at around 50 K, which is where the heating and cooling curves merge. In Figure 1(c) we plot R(T) below 50 K for various LED fluences. With increasing fluence, the resistance decreases until it saturates at our maximum power density, which is also evident in the normalized resistance as a function of power density and in the current-voltage curves shown in the inset of Supplemental Material for Figure S2 power density dependence. The non-volatile resistance changes at low temperature described next is another key aspect of our work. Figure 2 displays the photoexcited relaxation dynamics of the PNO resistance in three-time regimes: when the light is off, $0 \leq t < t_{ON}$; while we shine light on the sample $t_{ON} \leq t < t_{OFF}$; and after we turn off the light $t \geq t_{OFF}$, with $t_{ON}$ = 81 s, $t_{OFF}$ = 185 s. We measured the response up to 8 hours after $t_{OFF}$. At lowest temperature, resistance drops quickly at $t_{ON}$ to a value of 12% of the original resistance ($\Delta R_{ON}$), but only recovers to about 40% of its initial state after 8 hrs ($\Delta R_{OFF}$). The only way to recover the full state of resistance is warming up the sample above the transition and cycling it back down. As the temperature increases, the light-induced resistance drops as a percentage of its initial resistance decreases, but the recovery ratio between $\Delta R_{OFF}$ and $\Delta R_{ON}$ increases. At 50 K, the transient resistance approaches the initial resistance quickly after the light is off and the effect becomes volatile. This is represented in Figure 2c, where we plot the recovery percentage, which can be thought of as the volatility, as a function of temperature. The light induced volatility is highest at high temperatures and becomes more non-volatile as the temperature is lowered.

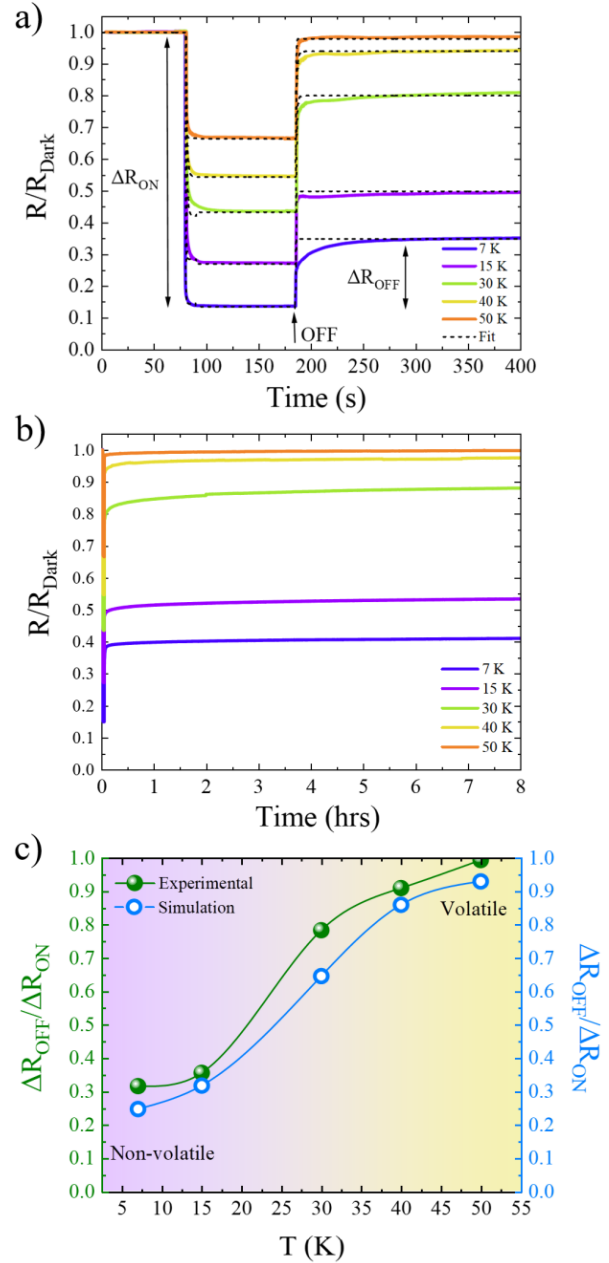

**FIG. 2.** (a) Time dependence of the resistance normalized to the initial insulating state before illumination taken at different temperatures: 7 K, 15 K, 30 K, 40 K and 50 K. At $t = t_{ON} = 81$ s, the sample is illuminated, and the light is turned off at $t_{OFF}=185$s. The dotted black line is the fit from our model. (b) The resistance curves plotted over the span of 8 hours. (c) The ratio of resistance percentage drop when the light is turned on over the recovery percentage after the light is turned off, plotted as a function of temperature for the experimental data (blue) and the model (red).

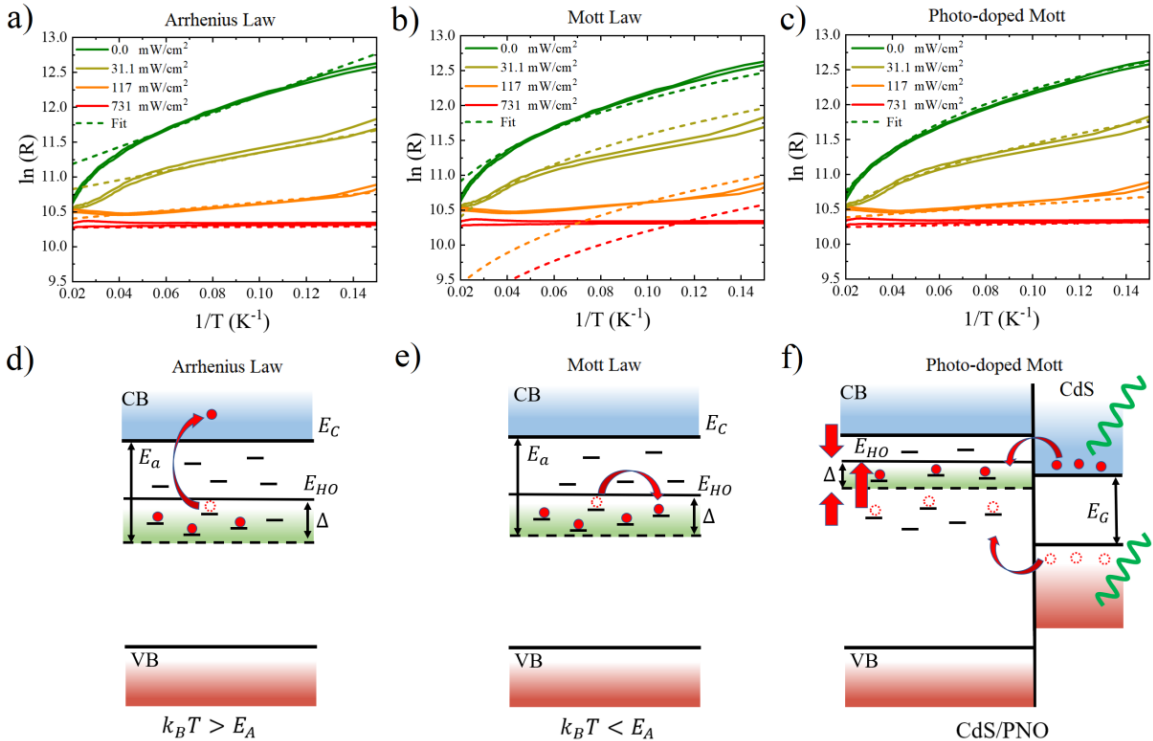

**FIG. 3.** The experimental R(T) curves at various light power densities plotted as ln(R) as a function of 1/T for comparison to: (a) the activated Arrhenius power law, (b) the Mott law, and (c) our phenomenological model that combines the previous two mechanisms. In the lower row, band diagrams for PNO pictorially show the effect of: (d) thermal activation, (e) Mott hopping mechanism, and (f) the photo-induced changes upon carriers entering the PNO from the CdS, shown as an adjacent band diagram. Here $E_{HO}$ is the highest occupied energy level, $E_A$ is the activation gap, CB is the conduction band, $k_B$ is Boltzmann's constant, and $E_G$ is the gap of CdS. Red filled dots are electrons and dotted empty dots are holes.

To interpret our data, we consider different models that must explain three aspects of our experimental observations (see Figure 3). These aspects are: (i) the intrinsic exponential R(T) behavior of PNO below the hysteresis regime, i.e. without the effect of light, (ii) the changes in the R(T) upon illumination, when charge carriers migrate across the interface between CdS and PNO, and (iii) the temporal dependence of the switching behavior. In addition, we consider in the R(T) behavior of pure PNO which has subtle differences compared to the heterostructure above 10K. We also incorporate the layered architecture of the device by including the contribution of the CdS-PNO interface and PNO resistance in a parallel configuration. The pure CdS resistance is not considered because it is one order of magnitude higher than PNO. A comparison between the pure PNO layer and CdS-PNO without light confirms the role of the CdS-PNO interface resistance even in the absence of light (see Supplemental Material for Figure S4 resistance of pure PNO).

First, we attempt to fit the data with and without light on a bare PNO film, using an Arrhenius model of thermally activated transport across a gap. This yields an exponential dependence on temperature, parametrized by the activation gap $E_a$, written in logarithmic-scale as follows

$$\log(R) = \log(R_0) + \frac{E_a}{k_B T} , \qquad (1)$$

While this model does not yield a good fit for the intrinsic R(T), it fits increasingly better to the illuminated R(T) curves with increasing power density (Fig. 3a). Thus, we postulate that activation physics play a role in the photoinduced transport, but an additional mechanism is involved. Next, we evaluate variable range hopping, i.e. "Mott law", to fit the intrinsic R(T) (Fig. 3b). In this model, at low temperatures, phonon-assisted tunneling of localized electrons leads to hopping conduction. As a consequence, a constant density of states close to the highest occupied energy level $E_{HO}$ produces a temperature dependent hopping length, hence called a variable range hopping (VRH). Inspired by this model we conjecture that the conduction results from hopping between shallow impurity states close to the conduction band edge $(E_{HO} - \Delta, E_{HO})$, with $\Delta$ the band width below $E_{HO}$, see Figure 3e. The logarithmic resistance is given by[50, 51]

$$\log(R) = \log(R_0) + \left(\frac{T_{Mott}}{T}\right)^{1/4}, \qquad (2)$$

with $T_{Mott} = \frac{\beta}{k_B g(E_{HO}) a^3}$, also referred to as the "Mott temperature" (for more details, see Supplemental Material for Conduction models). Intuitively, $T_{Mott}$ is given by the hopping energy scale[51]. In order to avoid assumptions on the fitting parameters' dependence on increasing power, we assume that the Mott temperature is independent of light power. In this case, the fit is much better for the intrinsic R(T), but starts to deviate substantially as the light power is increased (Fig. 3b). Thus, the transport in the low-power regime of our measurements is mostly governed by hopping, while the high-power regime is dominated by thermal activation. When the temperature increases, we find that the Arrhenius fit degrades while the Mott law improves, suggesting that the hopping energy is greater than the thermal activation energy. For completeness, we checked different VRH models like Efros-Shklovskii and d-dimensional[52], but the Mott law is the best fitting model (see Supplemental Material for Figure S7 conduction model fitting comparison).

Inspired by the previous observations, we introduce a phenomenological model that connects continuously the intrinsic and highly photodoped conductance regimes at all temperatures. We propose that the resistivity is given by

$$\log(R) = \log(R_0) + \frac{\Delta(\Gamma) - E_a}{k_B T} + \left(\frac{T_{Mott}(\Gamma)}{T}\right)^{1/4}. \qquad (3)$$

The previous expression results from combining Eq. (1) and Eq. (2) except for the inclusion the band width $\Delta$ which arises from considering the CdS-PNO interface effect without light (see Supplementary). Therefore, Eq. (3) accounts the whole range of light intensity used in our experiments. The essential feature of this model is that the band width $\Delta$ over which hopping occurs and $T_{Mott}$ both depend on the power density $\Gamma$, as follows. As the photoinduced carriers entering the nickelate layer from the CdS fill up the states near the PNO $E_{HO}$, there are less impurity levels available near $E_{HO}$. This effectively reduces the bandwidth over which the VRH model requires the Fermi surface to have a constant density of states (see the red arrows in Fig. 3f). Thus, the effective activation energy is $\Delta(\Gamma) - E_a$.

The result of these fits is shown in Figure 3c, revealing excellent agreement at all values of the power density across all temperatures. The values of the fitting parameters are shown in Table 1, showing good agreement with previous works[35, 36, 53, 54]. Further, a quantitative comparison between different fitting models corroborated that our model exhibits the lowest mean squared error. Now, we expand the model to study the relaxation process of the photoexcited PNO transport, where the time dependence appears naturally without any adjustable parameters. First, we consider the role of the density of charge carriers. In thermal equilibrium, the density of carriers is given by $N = n_{HO} e^{-(E_C - E_{HO})/k_B T}$, where $E_C$ is the conduction band energy, and $n_{HO}$ the carrier density[55]. Using that $E_C - E_{HO} = E_a - \Delta$, see Figure 3f, and the approximation $E_C - E_{HO} < k_B T$, we obtain a normalized carrier density given by

$$n = \frac{N - n_{HO}}{n_{HO}} = \frac{\Delta - E_a}{k_B T}. \tag{4}$$

Then, the resistance depends on the carrier density $n$ as follows:

$$\log(R) = \log(R_0) + n + \left(\frac{T_{Mott}(\Gamma)}{T}\right)^{1/4}, \tag{5}$$

where a continuity condition governs the temporal dynamics of $n$,

$$\frac{dn}{dt} = \frac{n_i(\Gamma) - n}{\tau}, \tag{6}$$

where $n_i(\Gamma)$ is the initial PNO carrier concentration and $\tau$ the relaxation time (more details in the Supplemental Material for Relaxation section Eq. (S7)). The resulting dynamics of the normalized resistance are plotted in Figure 2 as dotted lines, showing remarkable agreement at the onset of the photoinduced switching, and good agreement when the light is turned off. Importantly, our model does not require any additional parameters to yield these fits. In addition, it captures the photoinduced effect's varying volatility as a function of temperature very well (Fig. 2c).

| Power Density $\Gamma(mW/cm^2)$ | $\log(R_0)$ | $E_a$ (meV) | $\Delta(\Gamma) - E_a$ (meV) | $k_B T_{Mott}(\Gamma)$ (meV) |
|---|---|---|---|---|
| 0.0 | 7.3 | 1.03 | −0.29 | 658.59 |
| 31.1 | 8.9 | 0.56 | 0.15 | 26.81 |
| 117 | 10.38 | 0.25 | 0.25 | $10^{-5}$ |
| 731 | 10.28 | 0.01 | 0.02 | $10^{-5}$ |

Table 1: Fitting parameters for the photodoped-Mott model.

The phenomenological model can then be interpreted as follows. Electronic conduction in systems with an inhomogeneous distribution of donor levels below the conduction band is dominated by thermal excitation and hopping. The former occurs as thermally activated electrons jump to the conduction band (yielding the Arrhenius law) provided that thermal energy surpasses the activation energy. The latter is due to the electron hopping from occupied to empty donor sites. Further assumptions on the distribution of donor states gives rise to a broad range of variable range hopping models. We focus on the Mott law, which assumes that electron conduction is due to hopping between localized and uncorrelated electrons lying in a narrow, constant density of states band around the Fermi level.

Typically, in doped semiconductors, these two mechanisms are mutually exclusive since the activation energy is much greater than hopping energy barrier. However, we found that in PNO the mechanisms of thermal activation and electron hopping coexist because in PNO the activation energy and the hopping energy scales are similar. Furthermore, they can simultaneously be modified by photodoping since the injection of carriers alters the width of the band and shifts the Fermi level itself. From this, we estimate a lower threshold of the penetration depth of these photo induced carriers, based on the idea that the transport is only feasible when the penetration depth approximates the hopping length which we estimate at 5 nm using intrinsic parameters of PNO (see Supplemental Material for Hopping length/penetration depth estimate). In other words, since impurity state disorder restricts the conduction of carriers that manage to reach impurity sites, only when the dopants penetrate beyond 5 nm will they activate the transport via hopping. At low power levels (low photoinduced carrier density), the activation gap is much higher that the low (< 10 K) temperature kT impeding activated transport. Thus, in this regime the transport follows the Mott law. For T > 10 K, the activated and hopping mechanisms coexist. We also note that our model of the intrinsic R(T) at higher temperatures improves on the Mott law fit, comparing Figures 3b and

3c at low values of 1/T. The Arrhenius laws becomes more relevant at higher temperatures. Also, at high power values (high photoinduced carrier density), $E_A$ is much lower than kT throughout the whole insulating temperature regime, thus governing the activated transport. In the intermediate regimes of temperature and light power, both mechanisms contribute. As it pertains to the time relaxation, the volatility increases with temperature because the thermally activated transport is more volatile than the band structure changes that occur in the hopping regime. The changes in the band structure are more difficult to erase once the impurity states have been filled, and thus the behavior is non-volatile when governed by hopping. Since the fit parameters obtained in our model are in good agreement with other similar works[35, 36, 53, 54], we can conclude that the model based on the competition between thermal activation and variable range hopping is universal among nickelates.

In summary, we have developed an experimental approach to modify the resistive properties of a nickelate thin film by placing in proximity to a photoconductor such as CdS. This approach yields a reduction of the low-temperature resistance by a factor of 10 when the heterostructure is illuminated by white light. The effect is only visible below the onset of the first-order metal to insulator transition of the nickelate. At low temperatures, the changes are permanent, which may allow this concept to be used in technological applications that require the encoding of memory, such as in neuromorphic devices with synaptic plasticity based on hydrogenated nickelates[56]. The volatility of the resistive changes increases as the temperature increases. To explain these data, we use a model that combines two types of transport mechanisms: thermal activation and variable range hopping. Importantly, the model captures the resistance versus temperature dependencies very accurately below the transition, as well as the relaxation dynamics upon switching the light on and off without any further adjustable parameters. In addition, the photoinduced method presented here may also be applied to other oxides, to develop new useful optoelectronic functionalities.

**Acknowledgements**

This material is based upon work supported by the National Science Foundation under Grant No. DMR-2145080 and the Air Force Office of Scientific Research under award number FA9550-20-1-0242 (HN, NMV, PL, IKS). This material is based upon work supported by the Air Force Office of Scientific Research under award number FA9550-22-1-0318. A.F. was supported by the Research Corporation for Science Advancement via the Cottrell Scholar Award (27551) and the CIFAR Azrieli Global Scholars program.

**Competing interests**

The authors declare no competing interests.

**References**

[1] E. Dagotto, Complexity in Strongly Correlated Electronic Systems, Science, 309, 257-262 (2005).

[2] K.H. Ahn, T. Lookman, A.R. Bishop, Strain-induced metal–insulator phase coexistence in perovskite manganites, Nature, 428, 401-404 (2004).

[3] N. Vardi, E. Anouchi, T. Yamin, S. Middey, M. Kareev, J. Chakhalian, Y. Dubi, A. Sharoni, Ramp-Reversal Memory and Phase-Boundary Scarring in Transition Metal Oxides, Advanced Materials, 29, 1605029 (2017).

[4] Y. Tokura, N. Nagaosa, Orbital Physics in Transition-Metal Oxides, Science, 288, 462-468 (2000).

[5] J. Zaanen, G.A. Sawatzky, J.W. Allen, Band gaps and electronic structure of transition-metal compounds, Physical Review Letters, 55, 418-421 (1985).

[6] Y. Tokura, Correlated-Electron Physics in Transition-Metal Oxides, Physics Today, 56, 50-55 (2003).

[7] B. Keimer, S.A. Kivelson, M.R. Norman, S. Uchida, J. Zaanen, From quantum matter to high-temperature superconductivity in copper oxides, Nature, 518, 179-186 (2015).

[8] A.P. Ramirez, Colossal magnetoresistance, Journal of Physics: Condensed Matter, 9, 8171-8199 (1997).

[9] S. Middey, J. Chakhalian, P. Mahadevan, J.W. Freeland, A.J. Millis, D.D. Sarma, Physics of Ultrathin Films and Heterostructures of Rare-Earth Nickelates, Annual Review of Materials Research, 46, 305-334 (2016).

[10] R.M. Wentzcovitch, W.W. Schulz, P.B. Allen, VO2: Peierls or Mott-Hubbard? A view from band theory, Physical Review Letters, 72, 3389-3392 (1994).

[11] R. Rocco, J. del Valle, H. Navarro, P. Salev, I.K. Schuller, M. Rozenberg, Exponential Escape Rate of Filamentary Incubation in Mott Spiking Neurons, Physical Review Applied, 17, 024028 (2022).

[12] S. Cheng, M.-H. Lee, R. Tran, Y. Shi, X. Li, H. Navarro, C. Adda, Q. Meng, L.-Q. Chen, R.C. Dynes, S.P. Ong, I.K. Schuller, Y. Zhu, Inherent stochasticity during insulator-metal transition in $VO_2$, Proceedings of the National Academy of Sciences, 118, e2105895118 (2021).

[13] E. Qiu, P. Salev, L. Fratino, R. Rocco, H. Navarro, C. Adda, J. Li, M.-H. Lee, Y. Kalcheim, M. Rozenberg, I.K. Schuller, Stochasticity in the synchronization of strongly coupled spiking oscillators, Applied Physics Letters, 122, 094105 (2023).

[14] C. Ahn, A. Cavalleri, A. Georges, S. Ismail-Beigi, A.J. Millis, J.-M. Triscone, Designing and controlling the properties of transition metal oxide quantum materials, Nature Materials, 20, 1462-1468 (2021).

[15] D. Lee, B. Chung, Y. Shi, G.-Y. Kim, N. Campbell, F. Xue, K. Song, S.-Y. Choi, J.P. Podkaminer, T.H. Kim, P.J. Ryan, J.-W. Kim, T.R. Paudel, J.-H. Kang, J.W. Spinuzzi, D.A. Tenne, E.Y. Tsymbal, M.S. Rzchowski, L.Q. Chen, J. Lee, C.B. Eom, Isostructural metal-insulator transition in $VO_2$, Science, 362, 1037-1040 (2018).

[16] P. Homm, M. Menghini, J.W. Seo, S. Peters, J.-P. Locquet, Room temperature Mott metal–insulator transition in $V_2O_3$ compounds induced via strain-engineering, APL Materials, 9, 021116 (2021).

[17] E. Qiu, P. Salev, F. Torres, H. Navarro, R.C. Dynes, I.K. Schuller, Stochastic transition in synchronized spiking nanooscillators, Proceedings of the National Academy of Sciences, 120, e2303765120 (2023).

[18] R.S. Dhaka, T. Das, N.C. Plumb, Z. Ristic, W. Kong, C.E. Matt, N. Xu, K. Dolui, E. Razzoli, M. Medarde, L. Patthey, M. Shi, M. Radović, J. Mesot, Tuning the metal-insulator transition in $NdNiO_3$ heterostructures via Fermi surface instability and spin fluctuations, Physical Review B, 92, 035127 (2015).

[19] S. Lee, R. Chen, L. Balents, Landau Theory of Charge and Spin Ordering in the Nickelates, Physical Review Letters, 106, 016405 (2011).

[20] M. Medarde, C. Dallera, M. Grioni, B. Delley, F. Vernay, J. Mesot, M. Sikora, J.A. Alonso, M.J. Martínez-Lope, Charge disproportionation in $RNiO_3$ perovskites (R=rare earth) from high-resolution x-ray absorption spectroscopy, Physical Review B, 80, 245105 (2009).

[21] R.J. Green, M.W. Haverkort, G.A. Sawatzky, Bond disproportionation and dynamical charge fluctuations in the perovskite rare-earth nickelates, Physical Review B, 94, 195127 (2016).

[22] B. Lau, A.J. Millis, Theory of the Magnetic and Metal-Insulator Transitions in $RNiO_3$ Bulk and Layered Structures, Physical Review Letters, 110, 126404 (2013).

[23] H. Park, A.J. Millis, C.A. Marianetti, Site-Selective Mott Transition in Rare-Earth-Element Nickelates, Physical Review Letters, 109, 156402 (2012).

[24] M.H. Upton, Y. Choi, H. Park, J. Liu, D. Meyers, J. Chakhalian, S. Middey, J.-W. Kim, P.J. Ryan, Novel Electronic Behavior Driving $NdNiO_3$ Metal-Insulator Transition, Physical Review Letters, 115, 036401 (2015).

[25] M. Medarde, M.T. Fernández-Díaz, P. Lacorre, Long-range charge order in the low-temperature insulating phase of $PrNiO_3$, Physical Review B, 78, 212101 (2008).

[26] J.Y. Zhang, H. Kim, E. Mikheev, A.J. Hauser, S. Stemmer, Key role of lattice symmetry in the metal-insulator transition of $NdNiO_3$ films, Scientific Reports, 6, 23652 (2016).

[27] X.K. Lian, F. Chen, X.L. Tan, P.F. Chen, L.F. Wang, G.Y. Gao, S.W. Jin, W.B. Wu, Anisotropic-strain-controlled metal-insulator transition in epitaxial NdNiO3 films grown on orthorhombic $NdGaO_3$ substrates, Applied Physics Letters, 103, 172110 (2013).

[28] D. Preziosi, L. Lopez-Mir, X. Li, T. Cornelissen, J.H. Lee, F. Trier, K. Bouzehouane, S. Valencia, A. Gloter, A. Barthélémy, M. Bibes, Direct Mapping of Phase Separation across the Metal–Insulator Transition of NdNiO$_3$, Nano Letters, 18, 2226-2232 (2018).

[29] A.S. Disa, D.P. Kumah, J.H. Ngai, E.D. Specht, D.A. Arena, F.J. Walker, C.H. Ahn, Phase diagram of compressively strained nickelate thin films, APL Materials, 1, 032110 (2013).

[30] G. Mattoni, P. Zubko, F. Maccherozzi, A.J.H. van der Torren, D.B. Boltje, M. Hadjimichael, N. Manca, S. Catalano, M. Gibert, Y. Liu, J. Aarts, J.M. Triscone, S.S. Dhesi, A.D. Caviglia, Striped nanoscale phase separation at the metal–insulator transition of heteroepitaxial nickelates, Nature Communications, 7, 13141 (2016).

[31] J. del Valle, R. Rocco, C. Domínguez, J. Fowlie, S. Gariglio, M.J. Rozenberg, J.-M. Triscone, Dynamics of the electrically induced insulator-to-metal transition in rare-earth nickelates, Physical Review B, 104, 165141 (2021).

[32] E. Mikheev, A.J. Hauser, B. Himmetoglu, N.E. Moreno, A. Janotti, C.G. Van de Walle, S. Stemmer, Tuning bad metal and non-Fermi liquid behavior in a Mott material: Rare-earth nickelate thin films, Science Advances, 1, e1500797 (2015).

[33] A.J. Hauser, E. Mikheev, N.E. Moreno, T.A. Cain, J. Hwang, J.Y. Zhang, S. Stemmer, Temperature-dependence of the Hall coefficient of NdNiO$_3$ thin films, Applied Physics Letters, 103, 182105 (2013).

[34] J. Liu, M. Kargarian, M. Kareev, B. Gray, P.J. Ryan, A. Cruz, N. Tahir, Y.-D. Chuang, J. Guo, J.M. Rondinelli, J.W. Freeland, G.A. Fiete, J. Chakhalian, Heterointerface engineered electronic and magnetic phases of NdNiO$_3$ thin films, Nature Communications, 4, 2714 (2013).

[35] Q. Guo, S. Farokhipoor, C. Magén, F. Rivadulla, B. Noheda, Tunable resistivity exponents in the metallic phase of epitaxial nickelates, Nature Communications, 11, 2949 (2020).

[36] G. Catalan, R.M. Bowman, J.M. Gregg, Metal-insulator transitions in NdNiO$_3$ thin films, Physical Review B, 62, 7892-7900 (2000).

[37] R. Mallik, E.V. Sampathkumaran, J.A. Alonso, M.J. Martinez-Lope, Complex low-temperature transport behaviour of RNiO$_3$-type compounds, Journal of Physics: Condensed Matter, 10, 3969 (1998).

[38] A.M. Alsaqqa, S. Singh, S. Middey, M. Kareev, J. Chakhalian, G. Sambandamurthy, Phase coexistence and dynamical behavior in NdNiO$_3$ ultrathin films, Physical Review B, 95, 125132 (2017).

[39] H. Navarro, J.d. Valle, Y. Kalcheim, N.M. Vargas, C. Adda, M.-H. Lee, P. Lapa, A. Rivera-Calzada, I.A. Zaluzhnyy, E. Qiu, O. Shpyrko, M. Rozenberg, A. Frano, I.K. Schuller, A hybrid optoelectronic Mott insulator, Applied Physics Letters, 118, 141901 (2021).

[40] C. Adda, H. Navarro, J. Kaur, M.-H. Lee, C. Chen, M. Rozenberg, S.P. Ong, I.K. Schuller, An optoelectronic heterostructure for neuromorphic computing: CdS/V$_3$O$_5$, Applied Physics Letters, 121, 041901 (2022).

[41] H. Navarro, A.C. Basaran, F. Ajejas, L. Fratino, S. Bag, T.D. Wang, E. Qiu, V. Rouco, I. Tenreiro, F. Torres, A. Rivera-Calzada, J. Santamaria, M. Rozenberg, I.K. Schuller, Light-Induced Decoupling of Electronic and Magnetic Properties in Manganites, Physical Review Applied, 19, 044077 (2023).

[42] S. Catalano, M. Gibert, J. Fowlie, J. Íñiguez, J.M. Triscone, J. Kreisel, Rare-earth nickelatesRNiO3: thin films and heterostructures, Reports on Progress in Physics, 81, 046501 (2018).

[43] S.D. Ha, U. Vetter, J. Shi, S. Ramanathan, Electrostatic gating of metallic and insulating phases in $SmNiO_3$ ultrathin films, Applied Physics Letters, 102, 183102 (2013).

[44] R. Scherwitzl, P. Zubko, I.G. Lezama, S. Ono, A.F. Morpurgo, G. Catalan, J.-M. Triscone, Electric-Field Control of the Metal-Insulator Transition in Ultrathin $NdNiO_3$ Films, Advanced Materials, 22, 5517-5520 (2010).

[45] S. Asanuma, P.-H. Xiang, H. Yamada, H. Sato, I.H. Inoue, H. Akoh, A. Sawa, K. Ueno, H. Shimotani, H. Yuan, M. Kawasaki, Y. Iwasa, Tuning of the metal-insulator transition in electrolyte-gated $NdNiO_3$ thin films, Applied Physics Letters, 97, 142110 (2010).

[46] X. Granados, J. Fontcuberta, X. Obradors, J.B. Torrance, Metastable metallic state and hysteresis below the metal-insulator transition in $PrNiO_3$, Physical Review B, 46, 15683-15688 (1992).

[47] M. Hepting, M. Minola, A. Frano, G. Cristiani, G. Logvenov, E. Schierle, M. Wu, M. Bluschke, E. Weschke, H.U. Habermeier, E. Benckiser, M. Le Tacon, B. Keimer, Tunable Charge and Spin Order in $PrNiO_3$ Thin Films and Superlattices, Physical Review Letters, 113, 227206 (2014).

[48] M.A. Novojilov, O.Y. Gorbenko, I.E. Graboy, A.R. Kaul, H.W. Zandbergen, N.A. Babushkina, L.M. Belova, Perovskite rare-earth nickelates in the thin-film epitaxial state, Applied Physics Letters, 76, 2041-2043 (2000).

[49] M.S. Saleem, C. Song, F. Li, Y. Gu, X. Chen, G. Shi, Q. Li, X. Zhou, F. Pan, Light Tuning of the Resistance of $NdNiO_3$ Films With $CoFe_2O_4$ Capping, physica status solidi (RRL) – Rapid Research Letters, 12, 1800186 (2018).

[50] A.L.E. Boris I. Shklovskii, Electronic Properties of Doped Semiconductors, Springer Berlin, Heidelberg (1984).

[51] N.F. Mott, Conduction in glasses containing transition metal ions, Journal of Non-Crystalline Solids, 1, 1-17 (1968).

[52] A.L. Efros, B.I. Shklovskii, Coulomb gap and low temperature conductivity of disordered systems, Journal of Physics C: Solid State Physics, 8, L49-L51 (1975).

[53] S. Harisankar, K. Soni, E. Yadav, K.R. Mavani, Strain-mediated effects of oxygen deficiency and variation in non-Fermi liquid behavior of epitaxial $PrNiO_{3-\delta}$ thin films, Journal of Physics: Condensed Matter, 31, 135601 (2019).

[54] K. Ramadoss, N. Mandal, X. Dai, Z. Wan, Y. Zhou, L. Rokhinson, Y.P. Chen, J. Hu, S. Ramanathan, Sign reversal of magnetoresistance in a perovskite nickelate by electron doping, Physical Review B, 94, 235124 (2016).

[55] D.A. Neamen, Semiconductor physics and devices : basic principles / Donald A. Neamen, 4th ed., international ed. ed., McGraw-Hill Higher Education, New York, (2012).

[56] H.-T. Zhang, T.J. Park, I.A. Zaluzhnyy, Q. Wang, S.N. Wadekar, S. Manna, R. Andrawis, P.O. Sprau, Y. Sun, Z. Zhang, C. Huang, H. Zhou, Z. Zhang, B. Narayanan, G. Srinivasan, N. Hua, E. Nazaretski, X. Huang, H. Yan, M. Ge, Y.S. Chu, M.J. Cherukara, M.V. Holt, M. Krishnamurthy, O.G. Shpyrko, S.K.R.S. Sankaranarayanan, A. Frano, K. Roy, S. Ramanathan, Perovskite neural trees, Nature Communications, 11, 2245 (2020).

# Supplemental Material

**Disentangling transport mechanisms in a correlated oxide by photoinduced charge injection**

Henry Navarro[1]*, Sarmistha Das[1]*, Felipe Torres[2,3]*, Rourav Basak[1], Erbin Qiu[1], Nicolas M. Vargas[1,4], Pavel Lapa[1,4], Ivan K. Schuller[1]† and Alex Frano[1]††

[1] Department of Physics, Center for Advanced Nanoscience, University of California, San Diego, 92093, USA.
[2] Department of Physics, Universidad de Chile, Santiago 7800024, Chile.
[3] Center for the Development of Nanoscience and Nanotechnology, CEDENNA, Santiago 9170124, Chile.
[4] General Atomics, PO Box 85608, San Diego, CA 92186, USA

The epitaxial PrNiO$_3$ (PNO) film on single crystal NdGaO$_3$ (NGO) ⟨101⟩ substrate using pulsed laser deposition (PLD) technique from a polycrystalline bulk PrNiO$_3$ target. The substrate temperature was t fixed at - 710 ºC during deposition while the O$_2$ pressure was maintained at constant value of– 300mTorr. A KrF excimer laser, with energy density of 2 J/cm$^2$, was used for the target ablation. After growth the film was annealed for 10 minutes at 1.5 Torr O$_2$ pressure at 710 ºC. The phase purity of PNO thin film was investigated using x-ray diffraction (XRD). The XRD data shows that the film is epitaxial under tensile strained state with no presence of impurity peaks. On top of PNO, an 80 nm thick CdS film was grown using rf magnetron sputtering from a CdS target, in a 2-mTorr pure argon atmosphere at a 180°C substrate temperature.

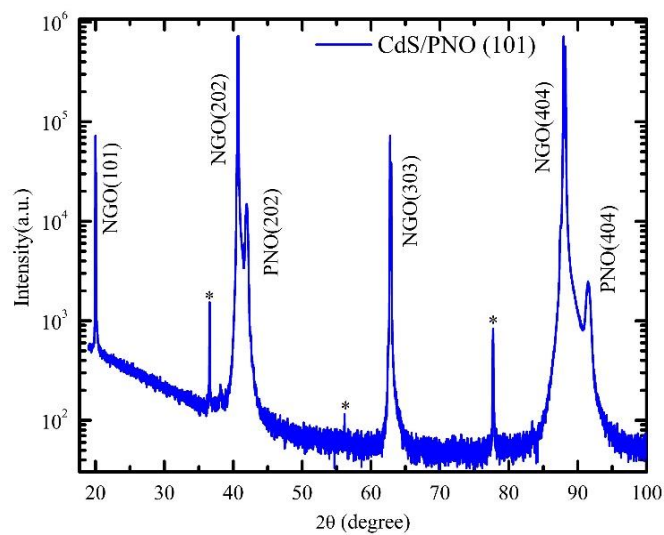

Figure S1: θ−2θ XRD patterns of PNO/NGO (101) film with a polycrystalline CdS. Asterisks (*) indicated peaks related to k$\beta$ peaks from the NGO substrate.

The normalized PNO insulating resistance at 7 K, decreases exponentially with increasing light power intensity. This is also shown in Figure 1c. The I(V) curve without light (green) clearly exhibits a non-ohmic behavior. As a function of light power density, the modified I(V) indicates the changes due to photodoping.

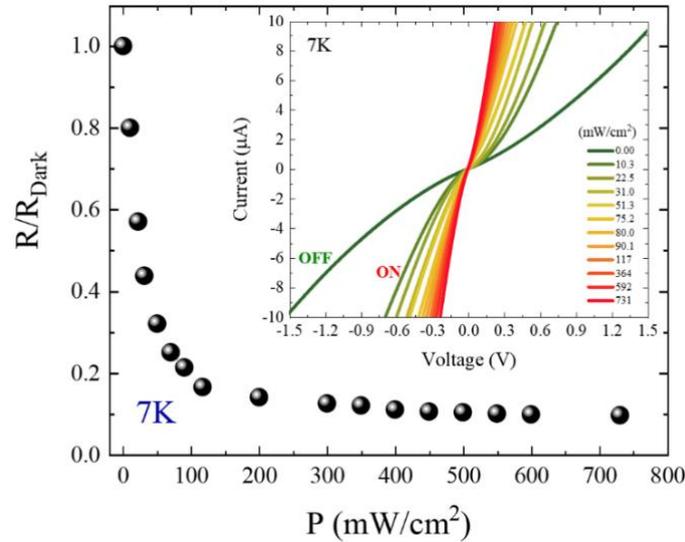

Figure S2: Power density dependence of the CdS/PrNiO$_3$ resistance at 7 K. The resistance is normalized to its value in the dark. The inset shows I(V) curves as a function of power density at 7 K.

We measured a bare CdS film (without the nickelate underlayer). When there is no light illumination, the resistance of the CdS film changes between $10^9$ to $10^{12}$ ohms between 300 and 200 K. Below 200 K it goes beyond our measurement limit (~tens of giga ohms). Importantly, this material is insulating at all temperatures and does not exhibit a metal-insulator transition. Furthermore, when 731 mW/cm$^2$ light power is applied to the CdS film, its resistance drops to ~$10^7$ ohms and does not exhibit any significant temperature dependence (see the Figure S3 below). On the other hand, the resistance of the PNO film is <$10^6$ ohms over a broad temperature range. The sample in our experiments (20 nm PNO sample) shows a 380 kilo ohms maximum resistance at 5 K. (Fig. 1c in the manuscript). Therefore, under light illumination, the lowest resistance of CdS is more than one order of magnitude higher than the highest resistance of PNO films in our experiments. As a consequence, the CdS layer cannot provide a parallel current path to the PNO film. This clearly reveals the CdS-PNO interface effect even in the absence of light. To prove this, we have grown a pristine PNO film (without CdS) and measured its resistance (Figure S4).

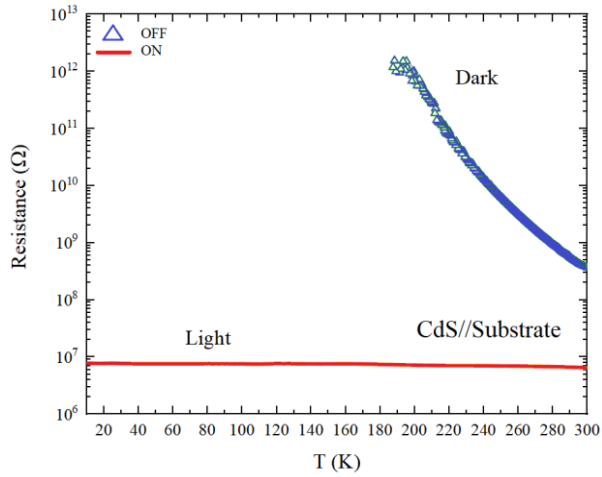

Figure S3: Electrical transport measurements as a function of temperature for pure CdS with and without light.

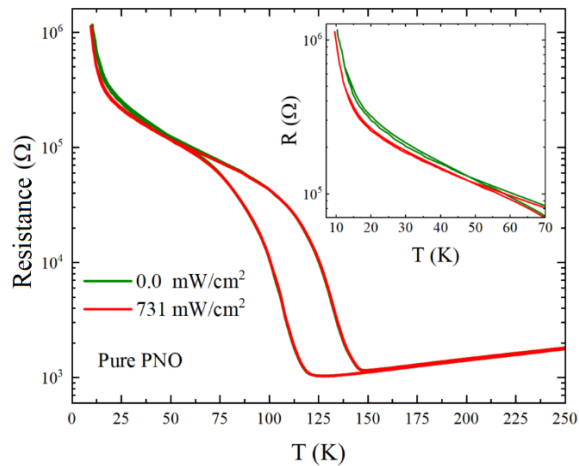

Figure S4: Resistance of pure PNO (20 nm) with light off (Green curve) and light on (red curve). Inset showed the insulating behavior with a 3K shift.

To check verify the reproducibility of these results, we fabricated an identical sample and obtained the curve below:

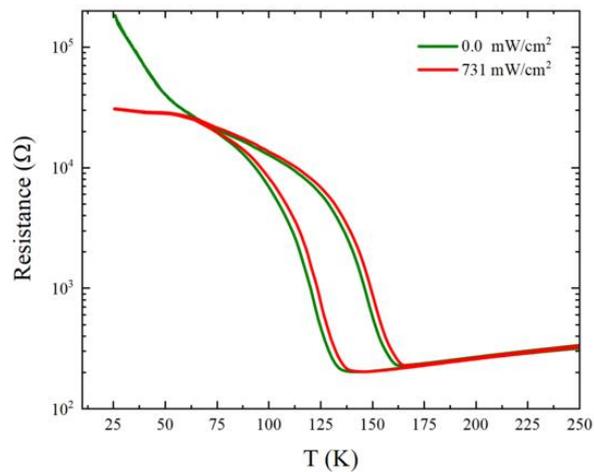

Figure S5: A experiment conducted on an identical CdS/PNO heterostructure, confirming the reproducibility of our results.

## Conduction Models

**Arrhenius law.**

According to the Arrhenius law, the low temperatures impurity effect and diffusion process yield the following temperature dependence of the electrical conductivity

$$\log(R) = \log(R_0) + E_a/k_B T, \qquad (S1)$$

where $E_a$ is the energy activation, and $k_B$ is the Boltzmann constant. The fitting parameters are

| Power Density (mW/cm²) | $R_0$ (Ω) | $E_a$ (meV) |
|---|---|---|
| 0.0 | $4.9 \times 10^4$ | 1.03 |
| 31.1 | $4.4 \times 10^4$ | 0.56 |
| 117 | $3.1 \times 10^4$ | 0.25 |
| 731 | $2.9 \times 10^4$ | 0.01 |

Table S1: Fitting parameters in Arrhenius law

Figure S5 shows that a highly-photodoped regime is governed by thermally activated electronic transport (Arrhenius), while the hopping mechanism (Mott) dominates the low-photodoped regime, since decreasing activation energy allows thermally activated electrons to access the conduction band. Notice that this prediction is only based on the activation energy fit.

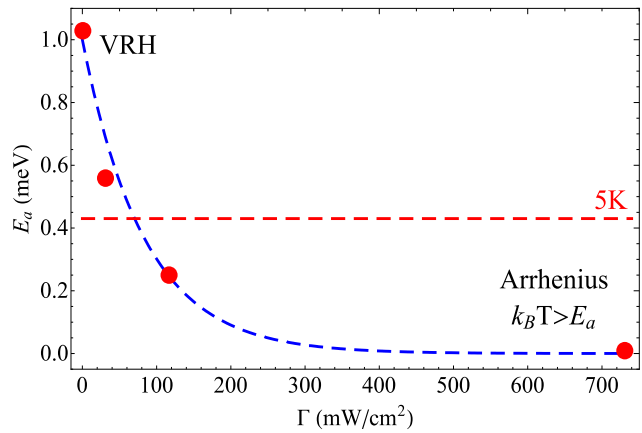

Figure S6: Light power-dependent activation energy. The red dashed line corresponds to thermal energy at 5K. As light power increases, activation energy decreases; above 100 mW/cm², thermally activated electrons can overcome the activation energy.

**Mott law.**

At low temperatures, phonon-assisted tunneling of localized electrons leads to hopping conduction where a constant density of states close to the highest occupied energy level produces a temperature dependent hopping length (Variable Range Hopping VRH). In this model the conduction results from the hopping between states concentrated in a narrow band near the highest occupied energy level $(E_{HO} - \Delta, E_{HO})$ [50,51], see Figure 3. The resistance is given by

$$\log(R) = \log(R_0) + \left(\frac{T_{Mott}}{T}\right)^{1/4}, \qquad (S2)$$

with $T_{Mott} = \beta/(k_B g(E_{HO})a^3)$. Using physical parameters given in the literature [50]: $E_{HO} = 0.4 eV$, $n_{HO} = ZN_A/V$ (the Avogadro number $N_A$), $Z = 4$, $V = 222 Å$, $g(E_{HO}) = 3n_{HO}/2E_{HO}$, with $\beta$ a constant $\beta \sim 10$, $a = 1\, nm$ [46], we obtain $n_{HO} = 1.8 \times 10^{22} cm^{-3}$, $g(E_{HO}) = 6.756\, cm^{-3} eV^{-1}$. We emphasize that we used parameters obtained by neutron-diffraction experiments [46, 55].

| Power Density mW/cm² | $R_0$ (Ω) | $T_{Mott}$ (K) |
|---|---|---|
| 0.0 | $5 \times 10^3$ | 1628 |
| 31.1 | $3 \times 10^3$ | 1628 |
| 117 | $1.1 \times 10^3$ | 1628 |
| 731 | $0.75 \times 10^4$ | 1628 |

Table S2: Fitting parameters in Mott law

**Photodoped-Mott.**

Inspired by previous results we introduce a phenomenological model that combines the essential features of the low and highly photodoped regime. Therefore, we propose that the resistivity is given by

$$\log(R) = \log(R_0) + \frac{\Delta(\Gamma) - E_a}{k_B T} + \left(\frac{T_{Mott}(\Gamma)}{T}\right)^{1/4}, \qquad (S3)$$

Figure 3 in the main text shows the comparison between the Arrhenius, Mott, and photodoped models. As light power density increases, the highest occupied energy level and energy band size (with constant density-of-states) decrease, allowing thermally activated transport.

**Relaxation.**

We use the aforementioned model to study the photoexcited relaxation dynamics of PNO. First, we consider the role of the density of charge carriers. In thermal equilibrium the density of carriers is given by [55]

$$N = n_{HO} e^{-(E_C - E_{HO})/k_B T}, \tag{S4}$$

where $E_C$ is the conduction band energy. Using that $E_C - E_{HO} = E_a - \Delta$ (see Figure 3f) and the approximation $E_C - E_{HO} < k_B T$, we obtain

$$n = \frac{N - n_{HO}}{n_{HO}} = \frac{\Delta - E_a}{k_B T}, \tag{S5}$$

Inserting (S5) into (S3), we get

$$\log(R) = \log(R_0) + n + \left(\frac{T_{Mott}(\Gamma)}{T}\right)^{1/4}, \tag{S6}$$

Photodoping caused by charge carriers' migration at the CdS/PNO interface can be simply described by considering the continuity equation,

$$\frac{dn}{dt} = \frac{n_i(\Gamma) - n}{\tau}, \tag{S7}$$

where $n_i(\Gamma)$ is the initial PNO carrier concentration and $\tau$ is the relaxation time. Solving equation (S7), we obtain $n(t) = n_i(\Gamma) + (n_f(\Gamma) - n_i(\Gamma))\exp(-t/\tau)$, $n_f(\Gamma)$ corresponds to the final concentration of carriers for a given final light power density. Therefore, the normalized resistance is

$$\frac{R(t)}{R(0)} = \frac{Exp\left(\log(R_0) + n(t) + \left(\frac{T_{Mott}(\Gamma)}{T}\right)^{1/4}\right)}{Exp\left(\log(R_0) + n(0) + \left(\frac{T_{Mott}(0)}{T}\right)^{1/4}\right)} = Exp\left(n(t) - n(0) + \left(\frac{T_{Mott}(\Gamma)}{T}\right)^{\frac{1}{4}} - \left(\frac{T_{Mott}(0)}{T}\right)^{1/4}\right) \tag{S8}$$

Since $T_{Mott}(0) > T_{Mott}(\Gamma)$ and $(e^{-t/\tau} - 1) = -t/\tau$, we have

$$\frac{R(t)}{R(0)} = Exp\left[-\left(T_{Mott}(0)/T\right)^{1/4}\right] Exp\left[-\left(n_f(\Gamma) - n_i(0)\right)\frac{t}{\tau}\right] \tag{S9}$$

Previous expression, Eq.(S8), describes the relaxation into two stage.

1) ON-stage: the sample is illuminated with $\Gamma = 731$ mW/cm² at constant temperature for 100s:

$$\frac{R(t)_{ON}}{R(0)_{ON}} = Exp\left[-\left(T_{Mott}(0)/T\right)^{1/4}\right] Exp\left[-\left(n_f^{ON}(\Gamma) - n_i^{ON}(0)\right)\frac{t}{\tau}\right] \quad (S10)$$

2) OFF-stage: the light is off for 8hrs:

$$\frac{R(t)_{OFF}}{R(0)_{OFF}} = Exp\left[-\left(T_{Mott}(\Gamma)/T\right)^{1/4}\right] Exp\left[-\left(n_i^{OFF}(\Gamma) - n_f^{OFF}(0)\right)\frac{t}{\tau}\right] \quad (S11)$$

We introduce superscripts to indicate the stage and subscripts to indicate the initial or final state of the stage. Therefore, the final state of the ON-stage corresponds to initial state of the OFF-stage, $n_i^{OFF}(\Gamma) = n_f^{ON}(\Gamma)$, we obtain

$$n_i^{OFF}(\Gamma) - n_f^{OFF}(0) = n_f^{ON}(\Gamma) - n_i^{ON}(0) + n_i^{ON}(0) - n_f^{OFF}(0) = \Delta n + \delta \quad (S12)$$

here $\delta = n_i^{ON}(0) - n_f^{OFF}(0)$ and $\Delta n = n_f^{ON}(\Gamma) - n_i^{ON}(0)$.

Thus

$$R(t)_{ON} = r_{ON} Exp\left[-\Delta n \frac{t}{\tau}\right] \quad (S13)$$

$$R(t)_{OFF} = r_{OFF} Exp\left[-(\delta + \Delta n)\frac{t}{\tau}\right] \quad (S14)$$

As a consequence, the normalized resistance

$$R(t) = \begin{cases} \frac{r_{ON} - R(t)_{ON}}{r_{ON} - R(0)_{ON}}, & 0 \le t \le T \\ R(T)_{ON} - R(t)_{OFF}, & T < t \end{cases} \quad (S15)$$

| T(K) | $r_{OFF}(\Omega)$ | $r_{ON}(\Omega)$ | $R(T)_{ON}(\Omega)$ | $\Delta n/\tau$ |
|---|---|---|---|---|
| 50 | 0.3 | 2 | 0.98 | 2.3 |
| 40 | 0.3 | 1.2 | 0.94 | 0.9 |
| 30 | 0.3 | 0.73 | 0.8 | 1 |
| 15 | 0.3 | 0.4 | 0.5 | 3.1 |
| 7 | 0.3 | 0.2 | 0.4 | 3.1 |

Table S3: Relaxation time parameters in photodoped-Mott.

**Hopping length/penetration depth estimate.**

We can estimate the hopping length/penetration depth of the photoexcited carriers $\lambda$ by using the Mott temperature definition as follows: $\lambda \sim (\beta/(T_{Mott} k_B g(E_{HO})))^{1/3}$, replacing $\beta \sim 10$, $g(E_{HO}) = 6.756 \ cm^{-3} eV^{-1}$, and $T_{Mott} \sim 10K$ we obtain $\lambda \sim 5nm$ as hopping length and penetration depth. Notice that we use the temperature where the thermal barrier created by the CdS/PNO affects the PNO behavior in the absence of light since photoinduced carriers are injected into the PNO if they overcome this energy.

**Efros-Shklovskii model.**

The derivation of the "photodoped Mott" as described above is based upon the assumption that the density of states near the highest occupied energy level is constant [52]. Including the role of the Coulomb interaction between localized electrons, Efros and Shklovskii demonstrated that the temperature dependence of the resistance is

$$\log(R) = \log(R_0) + \left(\frac{T_{ES}}{T}\right)^{1/2}, \quad (S16)$$

where $T_{ES} = \frac{e^2}{\kappa a k_B}$, $e$ is the electron charge, $\kappa$ is the dielectric constant and a is the localized length. Fitting parameters are

| Power Density (mW/cm²) | $R_0$ (Ω) | $T_{ES}$ (K) |
|---|---|---|
| 0.0 | $5.3 \times 10^4$ | 19.2 |
| 31.1 | $2.3 \times 10^4$ | 19.2 |
| 117 | $1.1 \times 10^4$ | 19.2 |
| 731 | $5.7 \times 10^3$ | 19.2 |

Table S4: Fitting parameters in Efros-Shklovskii model

**d-dimensional model.**

A natural generalization of the variable range hopping is given by

$$\log(R) = \log(R_0) + \left(\frac{T_d}{T}\right)^d \quad (S17)$$

with $0 < d \leq 1$. By introducing the mean squared error (MSE),

$$\text{MSE} = \frac{1}{N}\sum_{i=1}^{N}(y_i^{Model} - y_i^{Exp.})^2$$

$y_i^{Model}$ and $y_i^{Exp.}$, with i=1,..,N, stands for the theoretical prediction and experimental dataset respectively.

| Power Density | $R_0$ (Ω) | |
|---|---|---|
| mW/cm² | d =0.3 | d =0.4 |
| | $T_d = 481K$ | $T_d = 77K$ |
| 0.0 | $7.9 \times 10^3$ | $2 \times 10^4$ |
| 31.1 | $4.1 \times 10^3$ | $0.9 \times 10^4$ |
| 117 | $1.6 \times 10^3$ | $3.9 \times 10^3$ |
| 731 | $1 \times 10^3$ | $2.3 \times 10^3$ |

Table S5: Fitting parameters in the d-dimensional model.

| Power Density | $\text{MSE} = \frac{1}{N}\sum_{i=1}^{N}(y_i^{Model} - y_i^{Exp.})^2$ | | | | | |
|---|---|---|---|---|---|---|
| mW/cm² | Arrhenius $d = 1$ $T_d = \frac{E_a}{k_B}$ | Mott $d = 1/4$ $T_d = T_{Mott}$ | Efros-Shklovskii $d = 1/2$ $T_d = T_{ES}$ | d-dimensional $d = 0.3$ $T_d = 481K$ | d-dimensional $d = 0.4$ $T_d = 77K$ | Photo-doped Mott |
| 0.0 | 0.0334 | 0.0145 | 0.2440 | 0.0188 | 0.0485 | 0.0013 |
| 31.1 | 0.0358 | 0.0115 | 0.0121 | 0.0212 | 0.0275 | 0.0019 |
| 117 | 0.0039 | 0.4995 | 0.1386 | 0.5964 | 0.5131 | 0.0026 |
| 731 | 0.0025 | 0.7276 | 0.5850 | 0.9319 | 0.9572 | 0.0016 |

Table S6: Comparison of mean squared error in the different models.

| Power Density | $\chi^2 = \sum_{i=1}^{N}(y_i^{Model} - y_i^{Exp})^2 / y_i^{Exp}$ | | | | | |
|---|---|---|---|---|---|---|
| mW/cm² | Arrhenius $d = 1$ $T_d = \frac{E_a}{k_B}$ | Mott $d = 1/4$ $T_d = T_{Mott}$ | Efros-Shklovskii $d = 1/2$ $T_d = T_{ES}$ | d-dimensional $d = 0.3$ $T_d = 481K$ | d-dimensional $d = 0.4$ $T_d = 77K$ | Photo-doped Mott |
| 0.0 | 0.456351 | 0.0662038 | 1.46547 | 0.0893378 | 0.261046 | 0.00507886 |
| 31.1 | 0.277265 | 0.0447929 | 0.0880675 | 0.139554 | 0.0458789 | 0.009691 |
| 117 | 0.0296043 | 3.85376 | 1.07201 | 4.58283 | 3.92564 | 0.0268566 |
| 731 | 0.026738 | 6.60128 | 4.93683 | 8.25609 | 8.20006 | 0.0201882 |

Table S7: Comparison of Chi-square analysis in the different models

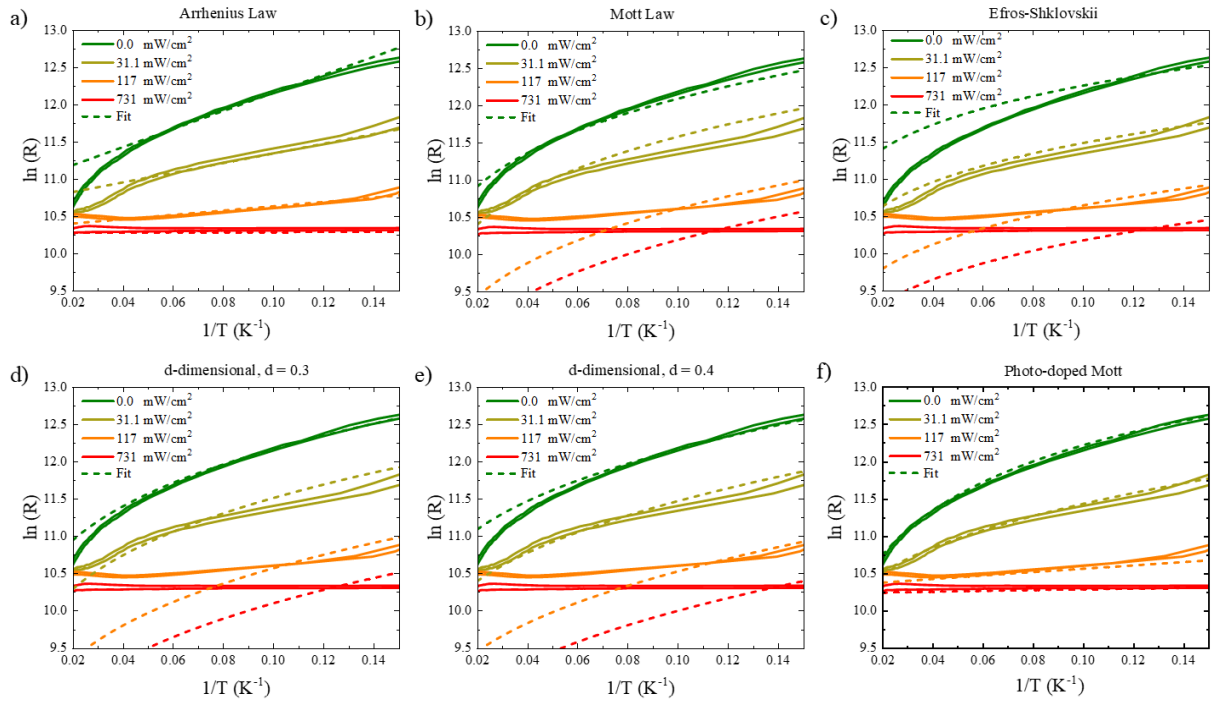

Figure S7: The conduction models fitting comparison. When we illuminate the device with low-power light, the variable range hopping models: Mott, Efros-Shklovskii, and d-dimensional, are in good agreement with the experiment, while Arrhenius law describes the low photodoped regime.

## Conduction mechanism in PNO

We contrast pure PNO (continous green curve) and CdS/PNO with no illumination (black continous curve) data, see Figure S8.

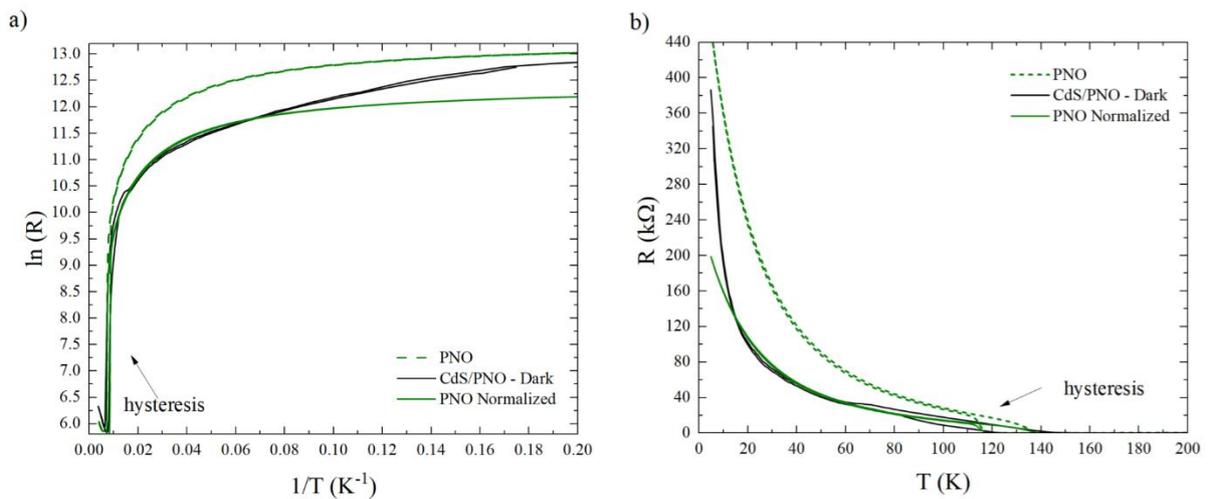

Figure S8: a) Experimental data pure PNO (green), CdS/PNO in dark (black), PNO normalized (dashed green). b) Experimental data in a lineal scale.

To compare two samples with different sizes, we introduce the PNO normalized data by adjusting the PNO resistance by $0.459 \times R_{PNO}$. The numerical factor accounts for the aspect ratio variation of the device (length/cross-sectional area). Below 10K, the normalized PNO and the CdS/PNO without light resistances differ significantly. This difference is attributed to the CdS-PNO interface in the absence of photodoping. As you seen below, our model is based on the simultaneous contribution of variable range hopping and thermally activation (VRHT). Now we show how to derive an analytical expression for logarithmic resistance based on this model. The total resistance of the heterostructure arises from the CdS, the interface (CdS/PNO), and the PNO resistances in a parallel configuration. Because the CdS resistance is orders of magnitude higher than the PNO resistance it does not contribute to the resistivity of the device. Therefore, the total resistance is written as

$$R_{Tot} = \frac{R_{PNO} R_{Int}}{R_{PNO} + R_{Int}}, \tag{S18}$$

$$\log(R_{Tot}) = \log(R_{PNO} R_{int}) - \log(R_{PNO} + R_{int}), \tag{S19}$$

In log-scale the exponential decay of $R_{PNO} + R_{int}$ is negligible as compared with $R_{PNO} R_{int}$, thus $\log(R_{Tot}) \approx \log(R_{PNO} R_{int})$, this aproximation was also checked numerically. Assuming that CdS-PNO interface conductivity is thermally activated across the energy barrier $E_{Int}$, the resistance of the interface is given by $R_{Int} \sim Exp(E_{int}/k_B T)$. Inserting this contribution in (S19) and using Arrhenius and Mott law we obtain

$$\log(R_{Tot}) = \log(R_0) + \frac{\Delta - E_a}{k_B T} + \left(\frac{T_{Mott}}{T}\right)^{1/4}, \tag{S20}$$

where $\Delta = \Delta_{OFF} + E_{int}$, $\Delta_{OFF}$ is the band width of pure PNO and $E_{int}$ is the thermal energy barrier of the CdS-PNO interface. Therefore, photodoping is given by the the variation of energy barrier of the interface.

Figure S9 displays the PNO normalized and CdS/PNO without light fitting data using distinct conduction models. The VRHT model best agrees with the experimental data of pure PNO and CdS/PNO. According to Table S7, in the absence of light, the gap between the conduction band and impurity states band is 64.5% lower due to the CdS/PNO interface. This interface effect does not include additional terms in equation (S18) but reduces the gap between the bands. Upon which we shine a light on the heterostructure photodoping mechanism is activated but the use the Equation (S20) to model the resistance; in this case, the VHRT has recalled as Photodoped Mott model.

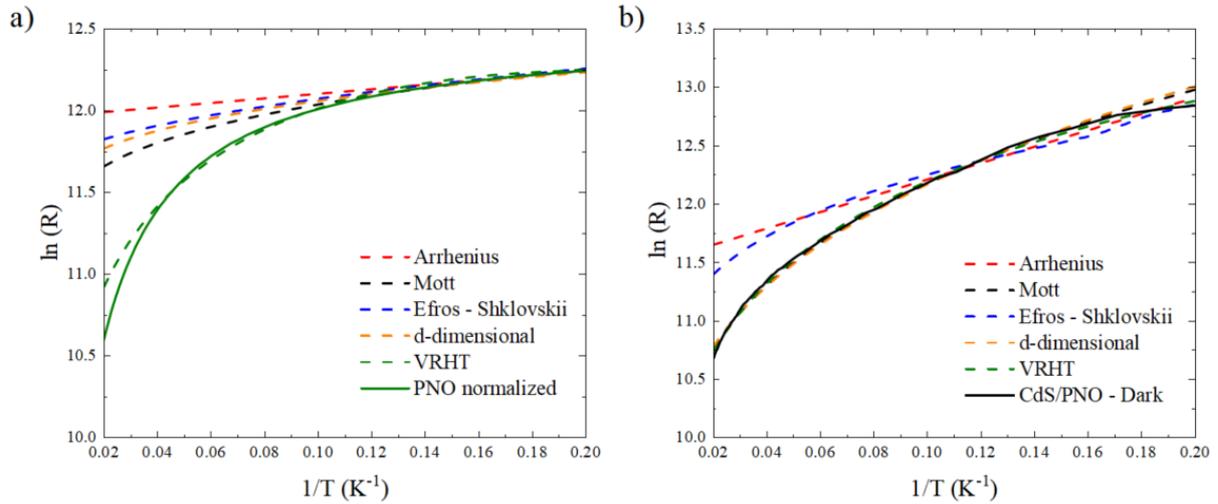

Figure S9: Conduction models fitting for pure PNO and CdS/PNO in dark.

| Parameters | PNO normalized | CdS/PNO - dark | %Difference |
|---|---|---|---|
| $R_0$ | 2220(Ω) | 1685(Ω) | 24.1 |
| $\Delta - E_a$ | -0.62(meV) | -0.22(meV) | 64.5 |
| $T_{Mott}$ | 6400 (K) | 6400 (K) | --- |

Table S8: Fitting parameters

**References**

[46] X. Granados, J. Fontcuberta, X. Obradors, J.B. Torrance, Metastable metallic state and hysteresis below the metal-insulator transition in PrNiO$_3$, Physical Review B, 46, 15683-15688 (1992).

[50] A.L.E. Boris I. Shklovskii, Electronic Properties of Doped Semiconductors, Springer Berlin, Heidelberg, (1984).

[51] N.F. Mott, Conduction in glasses containing transition metal ions, Journal of Non-Crystalline Solids, 1, 1-17 (1968).

[52] A.L. Efros, B.I. Shklovskii, Coulomb gap and low temperature conductivity of disordered systems, Journal of Physics C: Solid State Physics, 8, L49-L51 (1975).

[55] D.A. Neamen, Semiconductor physics and devices: basic principles / Donald A. Neamen, 4th ed., international ed. ed., McGraw-Hill Higher Education, New York, (2012).